\documentclass[12pt]{article}
\usepackage{latexsym,amsfonts,amsmath,amsthm,amssymb}

\title{No "NO-GO"}
\author{Andrei Khrennikov\footnote{International Center for Mathematical Modeling 
in Physics and Cognitive Sciences, Andrei.Khrennikov@msi.vxu.se; supported by EU-Network
 "QP and Applications''.} \\
MSI, University of V\"axj\"o, S-35195, Sweden}

\date{}
\begin{document}
\maketitle

\begin{abstract}In spite of the very common opinion we show that QM is {\bf not complete} and that it is possible to create
prequantum models providing finer description of physical reality than QM. There exists (at least 
in theoretical models) dispersion free states and the {\bf Heisenberg uncertainty principle can be 
violated.} Thus in spite of all ``No-Go'' theorems (e.g., von Neumann, Kochen and Specker,..., Bell) we found a {\bf realist 
basis} of quantum mechanics. I think that our model would totally satisfy A. Einstein who was sure
that QM is not complete and a finer description is possible.
\end{abstract}

Our representation is not standard model with hidden variables.
In particular, this is not a reduction of the quantum model to the classical one.
Moreover, we see that such a reduction is even in principle impossible. 
This impossibility is not the consequence of a mathematical theorem but it follows
from the physical structure of the model. By our model quantum states are very
{\bf rough images} of domains in the space of fundamental parameters - PRESPACE. Those
domains represent complexes of physical conditions. By our model both classical and quantum physics describe REDUCTION
of PRESPACE-INFORMATION.  As we have already mentioned, QM is not complete. In particular, there are 
prespace contexts which can be only  represented by a so called hyperbolic quantum model.

{\bf 1. ``Classical'' and ``quantum'' probabilities.}
The starting point of our considerations was the comparative analysis of so called 
{\bf classical and quantum probabilities.} Such an analysis was performed in my book [1]
and it demonstrated that, in fact,  there are no ``classical'' and ``quantum'' probabilities.
To understand the probabilistic foundations of QM in the right way we should forget about 
such funny (and totally meaningless) ideas as, e.g.,  {\bf irreducible quantum randomness}.
Instead of repeating  that quantum and classical (in some versions micro and macro)
worlds are described by totally different probabilistic calculi, we should find the conventional
probabilistic roots of so called ``quantum probabilities''. To find such roots we should start
with the most fundamental distinguishing feature of quantum probabilities, namely with 
interference of probabilities.

{\bf 2. Interference of probabilities.} There is the common opinion (e.g., see Feynman's book 
on path integrals) that there is crucial difference between classical and quantum
probabilistic rules for addition of probabilities of alternatives:
\begin{equation}
\label{F1}
P=P_1+P_2
\end{equation}
and
\begin{equation}
\label{F2}
P=P_1+P_2+2\sqrt{P_1P_2}\cos\theta.
\end{equation}
However, in 2001 I demonstrated that so called quantum rule for interference of probabilities 
can be easily derived, see, e.g. [2], by using the frequency von Mises approach to probability ---
probability as the limit of relative frequencies for a collective (random sequence).
Thus there is no contradiction between the classical frequency probability theory and quantum probability theory.
Quantum probability theory can be considered as a very special domain of the ``classical'' frequency probability theory.
Quantum probabilistic domain is {\bf a proper domain of ``classical'' frequency probability theory.}
It was found [2] that beside of quantum-like trigonometric interference (\ref{F2}) we can obtain 
{\bf hyperbolic interference:}
\begin{equation}
\label{F3}
P=P_1+P_2 \pm 2\sqrt{P_1P_2}\cosh\theta.
\end{equation}

{\bf 3. Contextual theory of probability.} Our derivation of interference of probabilities 
is based only on one postulate:

{\bf Contextuality of probabilities:} {\it Probabilities depend on complexes of physical conditions.}

We call complexes of physical conditions -- (physical) {\it contexts} and probability models 
based on this postulate -- {\it contextual probabilistic models.} 

In particular, the frequency probability model of von Mises is, in fact, a contextual probabilistic
model where contexts are represented by collectives. But R. von Mises did not pay attention
to the contextual aspect of his model and therefore he did not find formulas  (\ref{F2}), (\ref{F3}).

Recently [3] it was observed that by using the contextual interpretation of the conventional 
Kolmogorov model we can also obtain the interference of probabilities.

{\bf Conclusion:} {\it Interference of probabilities is not rigidly coupled with QM. 
Interference is a consequence of contextuality of probabilities.}

If you like we propose a probabilistic formalization of ideas of N. Bohr that the
whole experimental arrangement should be always taken into account. Unfortunately N. Bohr
never did this by himself. He was mainly concentrated on so called individual phenomena 
(see [4] on an extended analysis of views of N. Bohr),    i.e.,
{\it individual contextuality.} We are speaking about {\it statistical contextuality.}

{\bf 4. Contextually sensitive random variables.} Of course, nontrivial interference of probabilities can 
arise only for very special pairs $a, b$ of random variables, namely random variables which are
sensitive to perturbations of contexts produced by measurements. Let $C$ be a context. We call
the pair of (conventional, e.g. Kolmogorovian) random variables $a, b$ contextually sensitive
if the probability distribution for results of a measurement of the random variable
$b$ in the context $C$ differs from its probability distribution for results of a measurement
performed in a new context $C_a.$  The latter context is created by combination of 
in the original context $C$ and the context corresponding
to a measurement of $a$ (e.g., filtration with respect of values of $a).$ So the variable $b$ is sensitive 
to changes of a context $C$ induced by measurement of $a$ and vice versa. The $\cos \theta$ 
and $\cosh \theta$ in formulas (\ref{F2}) and (\ref{F3}) give the measure of statistical 
perturbation induced by perturbation of context.

We underline that we speak not about perturbations of physical systems induced by measurements
but by changes of complexes of physical conditions. Thus contextually sensitive random variables
need not be Heisenbergian.

It seems that N. Bohr in his famous reply to A. Einstein on the EPR-paradox tried to explain 
that physical observables under consideration are contextually sensitive. Unfortunately N. Bohr was 
sure that QM is complete.

{\bf 5. Quantum-like probabilistic models.} Since to get quantum-like interference of probabilities
we use only contextuality of probability and no ``really quantum features'' it would be natural 
to suppose that such an interference can be found not only in experiments with microsystems, but
in other experiments in which we measure contextually sensitive random variables. In [5] it was 
conjectured that quantum-like probabilistic behaviour might be observed in experiments with cognitive systems.
The main motivation was that cognitive systems are very complex information systems and mental observables 
are extremely sensitive to perturbations of context. This conjecture was confirmed by a series of psychological 
experiments [6]. Recently quantum-like behaviour of probabilities was also found in some games [7].

{\bf 6. Hilbert space representation of  a contextual Kolmogorov model.} 
Starting with the formula for interference of probabilities 
we constructed a Hilbert space representation of a contextual Kolmogorov model 
It was shown that for a Kolmogorov probability space 
$
{\cal K}=(\Omega, {\cal F}, {\bf P})
$
and a pair of {\it incompatible Kolmogorovian random variables} $b$ and $a$ we can construct a
natural quantum representation. This representation is rigidly based on a pair of
variables $b$ and $a$ --- fundamental  (for that
concrete representation of physical reality) observables. In particular, the standard quantum 
representation is based on the {\bf position and momentum observables.}

{\bf 7. Prespace.} Points of $\Omega$ are interpreted as {\it fundamental physical parameters}\footnote{If you like HV... 
But the general HV-approach was so discredited  by former investigations (since people wanted too much 
for such a HV-description) that we would not like even to refer to HV.}. 
We call $\Omega$ {\bf prespace} and fundamental parameters --- prepoints.

The main distinguishing feature of the representation map  is the huge 
{\bf compression of information.} In particular, every point represented in the conventional 
mathematical model of physical space by a vector $x \in {\bf R}^3$  
is the image of a subset 
$$
B_x=\{\omega \in \Omega:b(\omega)=x\}
$$ 
of $\Omega$ which can contain millions of prepoints. In the conventional quantum representation
of the prespace the fundamental variable $b=q$ is the {\bf position observable.}

{\bf 8. Classical probability and classical physics.} There is again a rather common tendency
to identify classical probability and classical physics. This is totally wrong idea. 

Classical physics provides the description of some physical phenomena by using Newtonian
(or Hamiltonian) mechanics and the model ${\bf R}^3$ for the ``physical space''
and ${\bf R}$ for the time, where ${\bf R}$ is the real line. So impossibility of classical
description of some phenomena should be considered as impossibility to embed such phenomena
in the Newtonian ${\bf R}^3$-model (with continuous real time). And if we read carefully
N. Bohr and W. Heisenberg we understand immediately that they had in mind precisely this 
idea on the impossibility of classical reduction of QM. However, later (mainly due to 
efforts of quantum logic and Bellian community) classical description was identified with
classical probabilistic description. This induces the above mentioned view:

{\bf classical probability iff classical physics.}

As we have already pointed out quantum probabilistic calculus can be obtained
in the classical (but contextual) probabilistic framework. Moreover, quantum 
probabilistic calculus can successfully used in some classical physical models [8].

However, the existence of a prequantum classical probabilistic model  does not
imply the reduction of quantum physics to classical. In our model it is impossible
to imbed quantum phenomena in Newtonian model in the physical space. There 
exist physical contexts which could not be described by using the physical 
space coordinates. Such contexts can be represented only by vectors of a Hilbert
space.

Thus our approach is totally in accordance with views of N. Bohr and W. Heisenberg.
However, I think that we should say "Farewell" to the physical ${\bf R}^3$-space 
and not to classical probability.

{\bf References}

1.  A. Yu. Khrennikov, {\it Interpretations of Probability.}
VSP Int. Sc. Publishers, Utrecht/Tokyo, 1999.

2.  A. Yu. Khrennikov, Linear representations of probabilistic transformations 
induced by context transitions. {\it J. Phys.A: Math. Gen.,} {\bf 34}, 9965-9981 (2001).
http://xxx.lanl.gov/abs/quant-ph/0105059.

A. Yu. Khrennikov, On foundations of quantum theory. 
Proc. Conf. {\it Quantum Theory: Reconsideration
of Foundations,} ed. A. Khrennikov. 
Ser. Math. Modelling,, {\bf 2}, 163-196,V\"axj\"o Univ. Press (2002).

 A. Yu. Khrennikov, What is really ``quantum'' in quantum theory? 
{\it Foundations of Probability and Physics-2,} 
Ser. Math. Modelling in Phys., Engin., and Cogn. Sc., vol. 5, 285-296,
V\"axj\"o Univ. Press, 2002.

3. A. Yu. Khrennikov, Classical and quantum spaces as rough images of the fundamental prespace.
http://xxx.lanl.gov/abs/quant-ph/0306069.

4. A. Plotnitsky, Quantum atomicity and quantum information: Bohr, Heisenberg, and quantum mechanics as
an information theory. Proc. Int. Conf. {\it Quantum Theory: Reconsideration
of Foundations.} Ser. Math. Modelling in Phys., Engin., and Cogn. Sc., vol 2.,
309-342, V\"axj\"o Univ. Press, 2002.

5. A. Khrennikov, {\it On cognitive experiments to test quantum-like behaviour 
of mind.} quant-ph/0205092 (2002).

6. E. Conte, O. Todarello, A. Federici, F. Vitiello, M. Lopane, A. Khrennikov,
A preliminar evidence of quantum-like behaviour in measurements of mental states.
quant-ph/0307201. 

7. A. Grib, A. Khrennikov, K.Starkov, Probability amplitude in quantum like games.
http://xxx.lanl.gov/abs/quant-ph/0308074.

6.A. Yu. Khrennikov, S. V. Kozyrev, Noncommutative probability in classical disordered systems.
{\it Physica A,} {\bf 326}, 456-463 (2003).

\end{document}